\documentstyle[12pt]{article}
\makeatletter
\newcommand{\singlespacing}{\let\CS=\@currsize\renewcommand{\baselinestretch}{1.0}\tiny\CS}
\newcommand{\doublespacing}{\let\CS=\@currsize\renewcommand{\baselinestretch}{1.5}\tiny\CS}

\begin{document}

\title{Some comments on bound state eigenvalues of PT-symmetric potentials} 

\author{S. Banerjee \&  R. Roychoudhury\thanks{{\bf E-mail : raj @ 
www.isical.ac.in}}\\ Physics \& Applied Mathematics Unit\\ Indian
Statistical Institute\\  203 B. T. Road\\ Kolkata 700035\\ INDIA} 

\date{}

\maketitle

\vspace*{0.4cm}

\centerline{{\bf ABSTRACT}}

\vspace*{0.3cm}

\thispagestyle{empty}

\setlength{\baselineskip}{18.5pt}

Using purely physical arguments, it is claimed that for ID Schr\"{o}dinger
operators with complex PT-symmetric potentials having a purely real,
attractive potential well and a purely imaginary repulsive part, bound state
eigenvalues will be discrete and real. This has been illustrated with several
potentials possessing similar properties.

\newpage

Complex potentials are encountered in a variety of diverse situations ranging
from conventional quantum mechanical scattering[1] problems to field theory,
population biology and quantum chemistry[2]. The solution of
Schr\"{o}dinger's equation with such complex potentials is complicated by the
fact that the hermiticity of the Hamiltonian is lost because it is not
invariant to parity (P) or time reversal (T). Inspired by the result obtained
by Bessis, Bender et al[3] suggested that the commutation of the product PT is
a possible mechanism of weakening the standard requirements of hermiticity, and
the so-called ``PT-symmetric'' quantum mechanics acquired revewed
interest[4-14].

For such Hamiltonians, the Schr\"{o}dinger operator

$$\displaystyle{H ~=~ \frac{d^2}{dx^2} + V(x)}
\eqno {(1)}$$
is PT invariant if the potential $V(x)$ satisfies 
$$\displaystyle{\left[ V(x) \right] ~=~ \left[ V(-x) \right]^*}
\eqno{(2)}$$

We consider a class of such complex potentials which can be represented in the
so called supersymmetric form [13]. 

$$\displaystyle{V^{(1), (2)} = u^2 \pm u^\prime} \eqno {(3)}$$
where $u$ is a complex function of $x$ and prime denotes differentiation $w$.
$r$. $t$. $x$. $u$ can be expressed explicitly as $a(x) + ib(x)$ where $a(x)$
and $b(x)$ are certain real, continuously differentiable functions in $R$. So
we have 

$$\displaystyle{ V^{(1), (2)} ~=~ \left( a^2 - b^2 \pm a^\prime
\right) + 2 iab \pm b^\prime} \eqno{(4)}$$

In earlier literature involving exactly solvable as well as numerical and
WKB[14] procedure based eigenvalues, one important feature of many complex
potentials has not been mentioned. Recently Ahmed[21] observed that for
certain asymptotically vanishing potentials all the eigen values are real when
the real part is stronger than the imaginary part.We wish to comment, sololy
on the basis of physical arguments, that, an inspection of these potentials
reveal that only real, bound state, discrete eigenvalues can be present due to
the repulsive nature of the imaginary part of the complex potential and
presence of a purely real potential well. We elaborate this point with a few
illustrative examples of the type given by Eqn.(3) and our arguments are valid
for harmonic and cubic as well as complex anharmonic oscillators. For the sake
of simplicity coupling constants are taken to be unity.

For complex potentials $\displaystyle{V(x) = V_R (x) - i V_I (x)}$ where
$V_R$, $V_I$ are real, the differential conservation relation for the position
probability density[15] is given by

$\displaystyle{P (x, t) = u^* (x, t) ~ u(x, t) = \left| u (x, t) \right|^2}$,
( $u$ being the wave function) and vector probability current density 

$$\begin{array}{l}
\displaystyle{S (x, t) \sim \left[ u^* \nabla u - (\nabla u^*) u \right]}\\
\displaystyle{ \frac{\partial P}{\partial t}~ (x, t) + \nabla \cdot
S (x, t) ~=~ - \frac{2 V_I}{\hbar} ~P (x, t)}\\
\end{array} \eqno{(5)}$$
since $P(x, t)$, is non-negative, $V_I \geq 0$ indicates that the R. H. S. of
eqn.(5) is a sink, whereas $V_I < 0$ is a source of probability. Therefore,
from this probability conservation constraint and for physical problems of
interest, like inelastic neutron scattering and absorption from the nucleus,
as discussed in[6], we assume that $V_I \geq 0$, i e the sign of imaginary
part of the complex potential is dictated by these physical requirements.

The first example that came to illustrate these remarks is a localized
potential belonging to the category described by eqn.[3]. Here the functions
$a(x)$ and $b(x)$ are given by $\displaystyle{a(x) = \frac{1}{x}}$ and
$\displaystyle{b(x) = \frac{\lambda}{x^2} ,}$ $\lambda$ being a real coupling
constant whose sign is determined by the considerations described in the
previous paragraph. The supersymmetric potential pair is constructed as
follows : Define $\displaystyle{ W^+ = a + ib}$ and $\displaystyle{W^- = a -
ib}$.  Then, 
$$\displaystyle{V^{(1)} ~=~ W^{+2} + W^{+ \prime} = -
\frac{\lambda^2}{x^4}} \eqno {(6a)}$$
and
$$\displaystyle{V^{(2)} ~=~ W^{-2} - W^{- \prime} = \frac{2}{x^2} - 
\frac{\lambda^2}{x^4} + \frac{4 i \lambda}{x^3}}  
\eqno {(6b)}$$
where $\lambda$ is positive, from preceding arguments. This pair of potentials
of eqn.(6) vanish at infinity faster than coulombic type potentials
$(\frac{1}{x})$ and are plotted in figs. $1(a-c)$ with figs. 1(a) and (c)
depicting the real parts of $V^{(1)}$ and $V^{(2)}$ and fig. 1(b) the
imaginary part of $V^{(2)}$. For this supersymmetric pair, the real parts
start out at $- \infty$ and constitute an attractive potential well below the
real axis before becoming vanishingly small at large distances from the
origin. On the other hand, for the imaginary part of the complex potential
$V^{(2)}$, (which is $\geq 0$, from the discussions of eqn.(5)), the potential
starts at $+ \infty$ at the origin and falls off to zero as the inverse cube
of the distance from the origin. This is a purely repulsive potential which
cannot contribute to bound state eigenvalues. So, for this supersymmetric
potential pair, the solution of Schr\"{o}dinger's equation for bound state
eigenvalues essentially constitutes the solution of eqn.(1) for real
potentials, thus preserving hermiticily and real discrete bound state
eigenvalues can be obtained in principle. 

The second and third supersymmetric pair of potentials to be discussed are non
localized and obey eqn (3). They are[16]

$$\displaystyle{w_1 ~=~ \frac{1}{x + i} - i (x + i)^2}  
\eqno {(7a)}$$

$$\displaystyle{w_2 ~=~ - \left[ \frac{1}{x-i} -i (x- i)^2 \right]}  
\eqno {(7b)}$$
and [9]
$$\displaystyle{V_1^- (x) ~=~ \frac{2}{(x+i)^2} - (x+i)^4}  
\eqno {(8a)}$$

$$\displaystyle{V_2^- (x) ~=~ - 4i (x-i) - (x-i)^4}  
\eqno {(8b)}$$

These two pairs of manifestly PT-symmetric potentials have been plotted in
figs.2(a-d) and 3(a-b) respectively, the second pair in fig.3 viz 3 and 3d
being almost identical is not plotted. Again, it found that the imaginary part
of such potentials do not form attractive potential wells and so cannot
contribute to the solution of Schr\"{o}dinger's equation for the bound state
problem, which reduces to a problem with a real potential well for the
Schr\"{o}dinger operator of eqn.(1).

Another PT-symmetric potential pair which obeys eqn.(3) is the supersymmetric
version of the modified Poeschl-Teller hole with coupling constant $\mu$ (and
parameters $\lambda$ and $\displaystyle{\tilde{\lambda} (\tilde{\lambda}-1) =
\frac{\lambda^2}{\mu^2} - \frac{1}{4}}$), which has also been solved
analytically[21] elsewhere, and which we include here for argument's sake 

$$\displaystyle{V_P^{(1)} ~=~ \frac{\mu^2}{4} - \mu^2 \left[ \tilde{\lambda}
(\tilde{\lambda} - 1) + 1 \right]~sech^2~ \mu x - 2 i \lambda \mu~sech~\mu x~
tanh~\mu x} \eqno {(9a)}$$

$$\displaystyle{V_P^{(2)} ~=~ \frac{\mu^2}{4} - \mu^2 \tilde{\lambda}
(\tilde{\lambda} - 1)~sech^2~ \mu x} \eqno {(9b)}$$

The first of these potentials are plotted in fig.4(a) (for the real part) and
fig.4(b) for the imaginary part. The real part constists of a bounded
potential well with discrete eigenvalues[13] whereas, the imaginary part is
a repulsive potential due to the conditions imposed by eqn.(5) (the coupling
constants $\mu$, $\lambda$ having the appropriate signs) and does not
contribute to the real discrete bound state eigenvalues.

For harmonic and anharmonic oscillator types of potentials with positive
coupling constants for the real harmonic and biharmonic components, no bound
state eigenvalues are produced although real, discrete eigenvalues extending
to continium have been obtained. For these harmonic oscillators, negative
coupling constants on the other hand would lead to discrete, bound states. 

For a potential of the type given in [17] viz., a real harmonic oscillaor
(with coupling constant $\mu \geq 0$) and a cubic oscillator with a purely
imaginary coupling constant $g$ viz.,
$\displaystyle{V(x) = \mu x^2 + i g x^3}$, where $g$ is necessarily positive
as a consequence of eqn.(5), it is well known that the real part (fig.5(a))
gives real, discrete, equidistant eigenvalues which extend to infinity. The
imaginary part, however, (fig.5(b)) does not possess any attractive potential
well and therefore cannot contribute to bound state, real eigenvalues, either.

Finally, for a complex anharmonic oscillator potential[1] $\displaystyle{V(x)
= ax^4 + bx^3 + cx^3 + dx}$, with $b = i \beta$ and $d = i \delta$ for
PT-symmetry, it was earlier asserted that the coupling constant has to be
negative[19] for bound states, but recent suggestions[14, 19] state that in
spite of the manifest non-hermiticity of the related Hamiltonian, the
procedure of quantization may be kept equally well defined at any sign of {\it
a}. Thus the real part of this anharmonic oscillator can have (fig.6 (a))
discrete, real, bound states, whereas the imaginary part (fig.6 (b)) which is
dominated by the $i \beta x^3$ term, is positive and does not have a potential
well and so bound states are not possible so, in conclusion it may be stated
that for non-hermition Hamiltonians, whose complex potentials obey PT-symmetry
expressed by eqn.(2), physical arguments support the presence of real bound
state eigenvalues obtained only from the real part of the potential, which
constitutes an attractive potential well, whereas the purely repulsive
imaginary part does not contribute to these bound states. This postulate is
illustrated with several potentials that have been plotted with the magnitudes
of the coupling constants taken to be unity, for simplicity and without loss
of generality. However, some important exceptions[18, 20] have been found for
which non-hermitian Hamiltonians $H \neq H^+$ support perfectly stable bound
states. Recently, the presence of such exceptional stable resonances for which
$Im~E=0$ ($E$ eigenvalue) has been a subject of intensive study. To conclude
we argue that PT-symmetric potentials having a purely real attractive
potential well and a purely imaginary repuslive part will have discrete and
real eigen values.

\section*{References}

\begin{enumerate}
\item[[1]] Teshbach H., Porter C. E. and Weisskopt V. F. : Phys. Rev. {\bf
96}, 448 (1954).
\item[[2]] D. R. Nelson and N. M. Shnerb : Phys. Rev. {\bf E 58}, 1383 (1998);
\item[] \ N. Hatano and D. R. Nelson : Phys. Rev. Lett. {\bf 77} (1996); Phys.
Rev. {B 56}, 8651 (1997).
\item[[3]] D. Bessis, unpublished (1992); C. M. Bender and S. Boettcher :
Phys. Rev. Lett. {\bf 80}, 5243 (1998);
\item[] \ C. M. Bender and S. Boettcher : J. Phys A. Math. Gen {\bf 31} L273
(1998); 
\item[] \ C. M. Bender and K. A. Millon : Phys. Rev. D {\bf 55} R3255 (1997).
\item[[4]] M. Znojil : Phys. Lett. A {\bf 259} (1999) and references therein.
\item[[5]] C. M. Bender, F. Cooper, P. N. Meisinger, V. M. Savage : Phys.
Lett. A {\bf 224}, 2529 (2000).
\item[[6]] M. Znojil : Phys. Lett. A {\bf 259}, 220 (2000).
\item[[7]] M. Znojil : J. Phys. A. Math. Gen. {\bf 33}, L61 (2000).
\item[] M. Znojil : J. Phys. A. Math. Gen. {\bf 33}, 456 (2000).
\item[[8]] A. Khare and B. P. Mandal : Phys. Lett. A {\bf 272}, 53 (2000).
\item[[9]] P. Dorey, C. Dunning and R. Tateo : J. Phys. A {\bf 34}, L39
(2001). 
\item[[10]] F. Cannata, M. Ioffe, R. Roychoudhury and P. Roy : Phys. Lett. A
{\bf 281}, 305 (2001). 
\item[[11]] B. Bagchi, F. Cannata and C. Quesne : Phys. Lett. A {\bf 269}, 79
(2000). 
\item[[12]] G. L\'{e}vai, F. Cannata and A. Ventura : J. Phys. A {\bf 34}, 839
(2001).
\item[[13]] B. Bagchi and R. Roychoudhury : J. Phys. A {\bf 33}, L1-L3 (2000).
\item[[14]] C. M. Bender, F. Cooper, P. N. Meisinger \& V. M. Savage : Phys.
Lett. A {\bf 259}, 224 (1999).
\item[[15]] I. I. Schiff : Quantum Mechanics, 3rd ed., Mc Graw Hill, Tokyo,
(1968). 
\item[[16]] M. Znojil, F. Cannata, B. Bagchi and R. Roychoudhury : Phys. Lett.
B {\bf 483}, 284 (2000).
\item[[17]] G. A. Mezincescu : J. Phys A. Math. Gen. {\bf 33}, 4911 (2000).
\item[[18]] W. E. Lamb and R. C. Relherford : Phys. Rev. {\bf 72}, 241 (1947).
\item[[19]] S. B. Cramption, D. Klepporer and N. F. Ramsey : Phys. Rev. Lett.
{\bf 78}, 4914 (1997).
\item[[20]] V. I. Kukulin, V. M. Krasnopol'sky, J. Hor\'{a}cek, Theory of
resonances : Principles and Applications, Kluwer, Dordrecht, (1989).
\item[[21]] Z. Ahmed : Phys. Letts. A {\bf 282}, 343 (2001).
\end{enumerate}

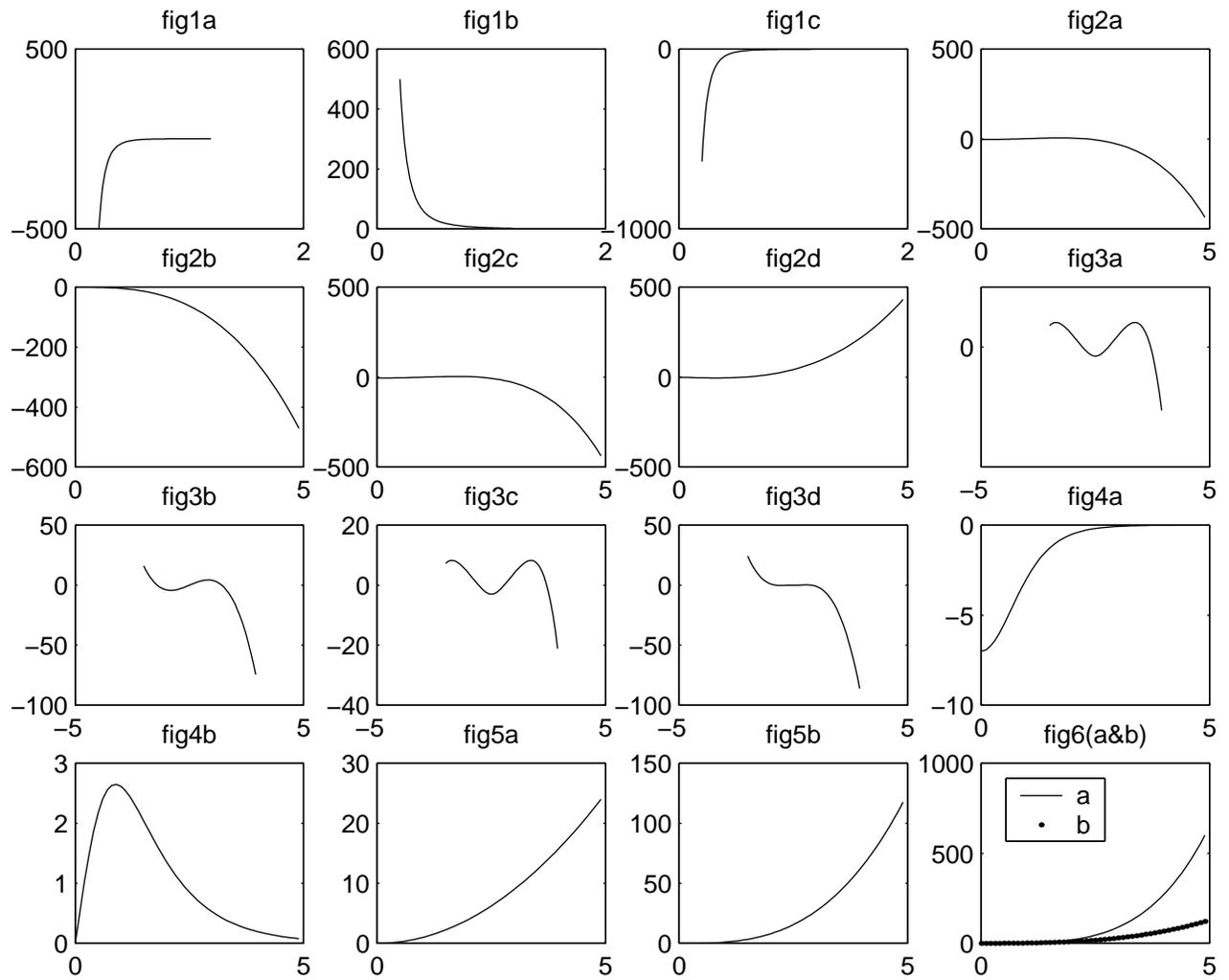
\end{document}